\documentstyle[aps,amssymb,12pt]{revtex}

\begin{document}
\author{Harald Hauglin}
\address{Department of Physics, University of Oslo, N-0316 Oslo, Norway}
\author{Nathan G. Woodard and Gregory P. Lafyatis}
\address{Department of Physics, The Ohio State University, Columbus, OH\\
43210-1106}
\title{Observation of hexatic liquid vortex matter in YBa$_{2}$Cu$_{2}$O$%
_{7-\delta }$}
\date{September 25, 2002.}
\maketitle

\begin{abstract}
An atomic beam probe is used to study the structure and dynamics of
quantized supercurrent vortex lines in YBa$_{2}$Cu$_{3}$O$_{7-\delta }$ at
temperatures up to within 0.7 K below $T_{C}$. Here we report the direct
observation of a vortex configuration with sample wide bond-orientational
order but only short range translational correlation. The data imply the
existence of an intermediate `hexatic' vortex line liquid phase. We find
that the hexatic liquid is in thermal equilibrium over a narrow temperature
range below $T_{C}$ and is quenched into an immobile hexatic glass at low
temperatures.
\end{abstract}


A magnetic field enters type II superconductors in the form of discrete
supercurrent vortex lines, each carrying a quantum of magnetic flux. The
microscopic structure and dynamics of vortex matter is not only crucial for
the application of superconductors - since it limits the capacity for
carrying electrical currents \cite{Brandt95} - but also has a fundamental
bearing on diverse topics such as the structure of liquid crystals \cite%
{Chou98} and the phase diagram of high density DNA mesophases \cite{Strey00}%
. In clean high-$T_{C}$ superconductors there is evidence for a first order
melting transition from a low temperature rigid and crystalline vortex
lattice to a high temperature mobile and disordered vortex liquid \cite%
{Cubitt93,Zeldov95,Roulin96,Schilling96}. For the case of YBa$_{2}$Cu$_{3}$O$%
_{7-d}$, there are indications that the melting transition changes character
at critical points both at high ($H_{\text{ucp}}$)\cite{Roulin98,Bouquet01}
as well as low magnetic fields ($H_{\text{lcp}}$)\cite{Willemin98,Paulius00}%
. The interesting region of the vortex matter phase diagram just below $T_{C}
$ has previously proved inaccessible to vortex-imaging-type studies and
there are few detailed results on the microscopic structure of the vortex
liquid.

In analogy with theories for 2D melting \cite{NHY}, it is predicted that
extended line objects such as nematic liquid crystals \cite{Toner83} and
quantized flux lines \cite{MarchettiNelson90} can exhibit a hexatic line
liquid phase with long range bond-orientational order but short range
translational correlation. The experimental determination of a hexatic phase
hinges on two issues: (i) the comparative range of translational and
bond-orientational order and (ii) whether the structure is in thermal
equilibrium. Early observations of hexatic correlations in the vortex array
by low temperature decoration techniques \cite{Murray90,Grier91} could not
resolve whether the images represented a frozen-in vestige of a hexatic
vortex liquid disordered due to thermal fluctuations or the ground state of
a lattice disordered by static random pinning defects \cite{Chudnovsky89},
such as found in charge density wave systems \cite{Dai92}. Decoration
studies of BSCCO samples indirectly indicated an equilibrium hexatic
structure significantly below $T_{C}$ \cite{Kim96}. Recent Hall probe
imaging of the vortex liquid phase in BSCCO \cite{Oral98} have been too
limited to provide data on the range of correlations of the vortex liquid.
Common to all these earlier reports is that they have probed a limited
number of vortices ($\lesssim 10^{4}$). We have used a new sensitive atomic
beam technique to study directly vortex correlations in a high quality YBa$%
_{2}$Cu$_{3}$O$_{7-\delta }$ single crystal in the low density regime below $%
H_{\text{lcp}}$ for temperatures to within 0.7 K below $T_{C}$. We have
probed correlations in a sample containing $3\times 10^{5}$ vortices, and
report evidence for a hexatic vortex line liquid with sample wide
bond-orientational order. We further demonstrate that the hexatic liquid is
in thermal equilibrium over a narrow temperature range below $T_{C}$ and is
quenched into an immobile hexatic glass at low temperatures.

Experimentally, to probe vortex correlations, we pass a beam of potassium ($%
^{39}$K) atoms near the sample and detect magnetic resonance (MR)
transitions driven by the microscopic magnetic field due to vortices (see
Fig. 1a). Transitions are resonantly driven in atoms when the atomic
velocity and the spacings of vortices along its path combine to make an
oscillating magnetic field component at the magnetic resonance frequency $%
f_{0}$. We actually measure the excitation probability for atoms as a
function of their velocity and work backwards to infer spatial
characteristics of the vortex array. If the atomic beam passes over an
ordered vortex lattice, the MR transition will be strongly driven for
velocities $v=df_{0}$ corresponding to vortex row spacings $d$ along the
beam (fig 1b). Formally, to first order, the velocity dependent MR
excitation probability, $P_{\text{MR}}(v)$, represents a projection of the
vortex matter structure factor \cite{Hauglin02}: 
\begin{equation}
P_{\text{MR}}(v)\thicksim \frac{1}{v^{2}}\int dq_{y}\left[ h({\bf q,}%
z,\lambda {\bf )}\right] ^{2}S_{2}({\bf q)}~,~q_{x}=2\pi f_{0}/v.
\end{equation}

Here $h({\bf q,}z,\lambda {\bf )}$ is the Fourier transform of the field
from a single vortex with penetration depth $\lambda $ at height $z$ above
the surface\cite{Marchetti92} and ${\bf q}=(q_{x},q_{y})$\ is the in-plane
wave vector. In analogy with x-ray scattering, $\left[ h({\bf q,}z,\lambda 
{\bf )}\right] ^{2}$ may be identified as the vortex {\em form factor}, and,
most importantly, $S_{2}({\bf q})=\int \exp (-i{\bf q\cdot r})g({\bf r}%
)d^{2}r$ \ is the 2D vortex lattice {\em structure factor} - the Fourier
transform of the vortex pair distribution function $g({\bf r})$ \cite{Ziman}%
. Details on the experimental method are published elsewhere \cite%
{Hauglin02,NbPapers}.

The experiments we report here were performed on a $0.7\times 1.7\times 0.1$
mm$^{3}$ YBa$_{2}$Cu$_{3}$O$_{7-\delta }$ single crystal ($T_{C}$ = 93.0 K, $%
\Delta T_{C}$ 
\mbox{$<$}
0.3 K). The sample was grown by a self-flux method \cite{Liang92} and
thermo-mechanically detwinned. Two electrical contacts ($\leq 0.1\Omega $)
are placed 1.3 mm apart on the top surface. The atomic beam skims the bottom
surface along the a-axis and probes a $0.7\times 1.0$ mm$^{2}$ area. A
typical MR scan takes 30 s to acquire.

Figure 2 shows MR profiles recorded during cooling at a rate 0.2 K/min in a
field 12 Oe parallel to the crystalline c-axis. We find that the MR profiles
show {\em two} broad peaks for all temperatures $T\leq 92.3$\ K for which we
detect a signal. Initially, upon cooling, there is a gradual screening of
the magnetic flux from the sample. Below 90 K, the flux density is constant
at 10.7 G, corresponding to a lattice constant $a_{0}\simeq 1.5$\ 
$\mu$%
m. This screening is seen as the slight increase of MR peak distances when
the sample is cooled. The two peaks represent (projected) vortex row
spacings for a vortex `lattice' with a unit cell that is misaligned with
respect to the atomic beam (similar to Fig. 1b, upper panel). In contrast,
either an isotropic vortex liquid or a `powder pattern' of randomly oriented
crystallites {\em without} sample wide orientational order would show a {\em %
single} bump representing the projection of a diffraction ring (Fig. 1b,
bottom panel)\cite{Hauglin02,NbPapers}. An immediate conclusion is that the
vortex array has sample wide orientational correlation in the entire
temperature range probed.

A detailed analysis of the MR signal is necessary in order to extract the
translational correlation length. The analysis is outlined here, details
will be published elsewhere\cite{Hauglin02}. We model the atomic excitation
using equation (1) and correct for saturation of the MR transition in a
rate-equation approach\cite{Hauglin02}. Saturation is important since it
broadens the MR peaks, especially when the magnetic field modulation is
strong: close to the sample surface and at low temperatures (short
penetration depth $\lambda $). We model the vortex pair distribution
function $g({\bf r})=\sum_{l,m}f\left( {\bf r-r}_{lm},\sigma \left(
r_{lm}\right) \right) $ , where $f({\bf r},\sigma )=(2\pi \sigma
^{2})^{-1}\exp (-r^{2}/2\sigma ^{2})$ and the sum runs over a lattice ${\bf r%
}_{lm}=l{\bf r}_{1}+m{\bf r}_{2}$\ spanned by ${\bf r}_{1}$ and ${\bf r}_{2}$
(execpt $l=m=0$). Orientational disorder is included by averaging the
structure factor $S_{2}({\bf q})$ for a range of primitive lattice vectors $%
{\bf r}_{1}$ and ${\bf r}_{2}$ rotated over an angular interval of width $%
\Delta \theta $. The computed transition probability (1) is corrected for
saturation and averaged over the relevant range $z<2$ 
$\mu$%
m for direct comparison to the experimental data. The main model freedom is
in the functional form of the displacement `dispersion' function $\sigma (r)$
and the orientational spread $\Delta \theta $. The unit cell parameters $%
{\bf r}_{1}$ and ${\bf r}_{2}$ are determined from the length scales for the
two MR peaks, their relative intensities, and the flux density $B,$ which is
measured independently using a miniature Hall array.

The best fit to the experimental line shapes (smooth curves in Fig. 2) is
for a triangular lattice that is tilted $18^{\circ }$ with respect to the
atom beam direction and compressed along that direction --- the crystalline $%
a$-axis of the sample\cite{Dolan89}. The fit is for a sample wide
orientational disorder $\Delta \theta \leq 2^{\circ }$, a dispersion
function of the form $\sigma (r)=\sigma _{0}(r/a_{0})^{p}$ with $\sigma
_{0}=(0.14\pm 0.01)a_{0}$ and roughness exponent $p=0.5\mp 0.1$. This
dispersion implies a liquid-type short range order. The data could not be
fitted using a quasi-long-range-order logarithmic dispersion predicted for
the Bragg-glass phase\cite{Giamarchi96}. The magnitude of the nearest
neighbor dispersion is consistent with the Lindeman criterion ($\sigma
_{0}/a_{0}\sim 0.1-0.15$) for a liquid near its melting point. Our
experimental data therefore shows a vortex array with a liquid-like
translational correlation length $\xi _{T}=(16\pm 8)a_{0}$ \ (determined by
the criterion $\sigma (\xi _{T})/a_{0}=0.5)$ that is much shorter than the
orientational correlation length $\xi _{\theta }\gg 500a_{0}$ (sample size).

Does this vortex pair distribution function represent a hexatic phase with
true long range bond-orientational order\cite{MarchettiNelson90} or does it
reflect finite size effects in a vortex liquid with slowly decaying, yet
short range, orientational order? Toner\cite{Toner91} argues that point
defects affect translational order more than orientational order, and
predicts that the correlation lengths should scale as $\xi _{\theta }=\xi
_{T}^{2}/a_{0}$ in the {\em absence} of orientational coupling with the
underlying crystal. In this scenario, we would expect $\xi _{\theta
}<10^{3}a_{0}$. We estimate a limit $\xi _{\theta }>10^{5}a_{0}$ for the
orientational correlation length by assuming that the angular disorder
increases with distance as rapidly as the translational disorder, i.e. $%
\Delta \theta (r)\sim (r/a_{0})^{0.5}$. Here $\xi _{\theta }$ is defined by $%
\Delta \theta (\xi _{\theta })=30^{\circ }$. Clearly, this scaling is not
satisfied -- the orientational order is long ranged. The structural data in
fig. 2 is therefore consistent with a {\em hexatic} vortex array with true
long range orientational order. The absence of scaling is evidence that long
range orientational order is induced by a coupling to the underlying crystal%
\cite{Chudnovsky91}. \ A further indication of this coupling is a preferred
direction for the vortex array with respect to the crystalline axes\cite%
{Chudnovsky91}: we find repeated field cooling runs always produced vortex
arrays with the same orientation.

Over which temperature range is the hexatic configuration in thermal
equilibrium? The experiment shown in Fig. 3 addresses this issue. First, we
observe that sample wide orientational order is destroyed by applying a
strong current to the field cooled vortex array at 82 K. This is seen as the
disappearance of the two distinct diffraction peaks for currents exceeding
220 mA (Fig. 3 a). (Note: 220 mA is almost an order of magnitude below the
critical current for the sample at this temperature.) The broad MR line
shape seen for large currents is consistent with a vortex array {\em without}
sample wide orientational order. The stability of the disordered vortex
array after the current is switched off (Fig. 3a, bottom curve) shows that
the transport current has introduced plastic deformations - the vortex array
is {\em not} in thermal equilibrium at 82 K. Second, the sample is
subsequently warmed and we observe the annealing of the disordered vortex
array above 91 K (Fig. 3b); Specifically, we see the reappearance of the two
distinct diffraction peaks - indicating long range bond-orientational order.
This shows that the hexatic structure is in a thermal equilibrium (liquid)
state over a $\sim 2$ K interval below $T_{C}$.

For temperatures below 91 K the structure of the hexatic liquid is quenched
into a hexatic glass. This happens when the dynamics of the vortex liquid
and hence the evolution toward crystallization is stopped as thermal
fluctuation energies become smaller than potential wells due to pinning
centers. This scenario is supported by several experimental observations: 
{\em (i)} The minimum pinning energy is comparable to thermal energies at
the glass formation temperature. The threshold current ($I_{\min }=120-220$
mA at 82 K) for the onset of change in the vortex array structure (Fig. 3a),
provides an estimate for the minimum vortex pinning force, $F_{\min }=\phi
_{0}I_{\min }/w$ $\simeq 5\times 10^{-13}$ N. Here $w=0.7$ mm is the width
of the sample. Assuming a quadratic pinning potential with a range
comparable to the coherence length ($\xi \sim 30$ \AA ), the corresponding
minimum threshold pinning energy is $U_{\min }/k_{B}\simeq F_{\min }\xi
/2k_{B}\sim 10^{2}K$. {\em (ii)} The amount of disorder does not change
measurably between 90 K and 10 K. The best fit model comparisons (blue
curves in Fig. 2) at 10 K and 91.5 K are for correlation functions $g({\bf r}%
)$ with identical amount of disorder. The effect of temperature is included
in the model line shapes only through the penetration depth dependence of
the vortex form factor $\left[ h({\bf q},z)\right] ^{2}$ \cite{Marchetti92}
in equation (1). The two curves are for $\lambda _{\text{ab}}=140$ nm and
400 nm at 10 K and 91.5 K respectively.

Theoretically, the hexatic flux liquid should be bounded by two phase
transitions\cite{NHY,MarchettiNelson90}. At high temperatures, it should
transform to an isotropic liquid by the proliferation of unbound
disclinations. At low temperatures, the hexatic liquid should undergo a
phase transition to a well ordered state (crystalline or Bragg glass\cite%
{Giamarchi96}) by freezing out unbound dislocations. Our experimental data
do not show any of these transitions. Assuming that the hexatic-to-isotropic
transition follows the established high field first order melting line\cite%
{Schilling96,Roulin96,Willemin98}, a simple extrapolation predicts $%
T_{m}\simeq 0.9998T_{C}$ at 12 G -- beyond the range of our method. On the
low temperature side, we have shown that crystallization is preempted by the
formation of a pinned hexatic vortex glass. The existence of a low density
hexatic vortex liquid near $T_{C}$ implies that the single first order
melting transition for intermediate fields separate into {\em two }%
transitions in the region below the lower critical point $H_{\text{lcp}}$.
The robustness of $H_{\text{lcp}}$ with respect to oxygen vacancies and weak
point defects due to electron irradiation\cite{Paulius00}, further suggests
that a low field hexatic liquid phase could be a generic feature of the
magnetic phase diagram of YBa$_{2}$Cu$_{3}$O$_{7-\delta }$ below $H_{\text{%
lcp}}$. We believe that hexatic order is imposed on the vortex liquid due a
coupling to the underlying electronic structure of the sample. The range of
existence of hexatic phases in other high-T$_{C}$ materials may therefore
depend on the anisotropy of the crystal structure.

\bigskip

\bigskip

FIG. 1. (a) Experimental principle: The time dependent magnetic field seen
by atoms moving near a sample with superconducting vortices drives magnetic
resonance (MR) transitions in atoms. If the field has a strong Fourier
component at the MR frequency $f_{0}$, then the transition $1\rightarrow 2$
will be strongly driven. For vortices spaced a distance $d$ apart, the MR
transition will be strongly driven for atoms moving with velocities near $%
v=df_{0}$ . (b) Model predictions for two different vortex lattices. In both
figures, the grey arrow indicates the direction of the atomic beam. For an
ordered vortex lattice, the transition probability of the atoms, , has
maxima at velocities corresponding to lattice row spacings along the beam.
The upper example shows a 10.5 G hexatic lattice rotated 15 degrees with
respect to the atomic beam - similar to the case reported in this work. The
two row spacings indicated, $d_{1}$ and $d_{2}$, are readily identified with
the two peaks in the data. Note: a triangular lattice with long range
translational order would give a similar signal but with much taller and
narrower peaks. For contrast, the lower example shows model predictions for
10.5 G isotropic liquid vortex structure. In reciprocal space, the signal
corresponds to a projection of 2D vortex structure factor (shown in the
insets) onto the beam direction.

\bigskip

FIG. 2. Atomic beam scans showing hexatic vortex matter structure. The MR
transition probability versus length scale is measured during field cooling
(at 0.2 K/min) in an applied field 12 Oe%
\mbox{$\vert$}%
\mbox{$\vert$}%
c. The blue curves are best fit model predictions for identical structure
factors with $\lambda _{ab}$= 140 and 400 nm at T=10 K and 91.5 K,
repectively. The presence of two distinct MR peaks indicate sample wide
orientational order. Note: the scale for each individual curve has been
adjusted to fit on the graph. The signal strength at 10 K are 10 times
larger than those at 91.5 K.

\bigskip

FIG. 3. MR profiles showing destruction of orientational order by a
transport current and subsequent restoration by annealing. (a) Current
dependence of MR profiles. The sample was initially field cooled to 82 K in
a 12 Oe field (top curve). MR transition profiles are shown for increasing
currents (middle three curves). The bottom curve shows the MR profile 1 hour
after turning off the 300 mA current. The broad bump at high current is due
to a vortex array without long range order. (b) Annealing of the disordered
vortex array. The top curve shows a MR profile due to the current induced
disordered vortex array (panel a) after the current was turned off. The
gradual appearance of two distinct peaks when annealing above 90 K shows the
re-appearance of long range bond-orientational order.

\end{document}